\documentclass[journal=nalefd,manuscript=article,layout=twocolumn,type=article]{achemso}

\usepackage[version=3]{mhchem} 

\newcommand{\BS}{Bi\textsubscript{2}Se\textsubscript{3}}

\newcommand{\ga}{$\alpha$}
\newcommand{\gb}{$\beta$}
\newcommand{\gbs}{$\beta^*$}
\newcommand{\aSb}{$\alpha$-Sb}
\newcommand{\bSb}{$\beta$-Sb}

\author{Conor Hogan}
\affiliation{Istituto di Struttura della Materia-CNR (ISM-CNR), Via del Fosso del Cavaliere, 00133 Roma, Italy
}%
\altaffiliation{
Dipartimento di Fisica, Universit{\`a} di Roma ``Tor Vergata'', Via della Ricerca Scientifica 1, 00133 Roma, Italy}
\author{Kris Holtgrewe}
\affiliation{Institut f\"ur Theoretische Physik and Center for Materials Research, Justus-Liebig-Universit\"at Gie{\ss}en, Heinrich-Buff-Ring 16, 35392 Gie{\ss}en, Germany}
\author{Fabio Ronci}
\affiliation{Istituto di Struttura della Materia-CNR (ISM-CNR), Via del Fosso del Cavaliere, 00133 Roma, Italy
}%
\author{Stefano Colonna}
\affiliation{Istituto di Struttura della Materia-CNR (ISM-CNR), Via del Fosso del Cavaliere, 00133 Roma, Italy
}%
\author{Simone Sanna}
\affiliation{Institut f\"ur Theoretische Physik and Center for Materials Research, Justus-Liebig-Universit\"at Gie{\ss}en, Heinrich-Buff-Ring 16, 35392 Gie{\ss}en, Germany}
\author{Paolo Moras}
\author{Polina Sheverdyaeva}
\author{Sanjoy Mahatha}
\affiliation{Istituto di Struttura della Materia-CNR (ISM-CNR), S.S. 14, km 163.5, I-34149 Trieste, Italy}
\author{Marco Papagno}
\affiliation{Dipartimento di Fisica, Universit\`{a} della Calabria, Via P.Bucci, 87036 Arcavacata di Rende (CS), Italy}
\author{Ziya S. Aliev}
\affiliation{Azerbaijan State Oil and Industry University, AZ1010 Baku, Azerbaijan}
\author{Mohammad B. Babanly}
\affiliation{Institute of Catalysis and Inorganic Chemistry, Azerbaijan National Academy of Science, AZ1143 Baku, Azerbaijan}
\author{Evgeni V. Chulkov}
\affiliation{Centro de F\'{i}sica de Materiales, CFM-MPC, Centro Mixto CSIC-UPV/EHU, Apdo. 1072, 20080 San Sebasti\'{a}n/Donostia, Basque Country, Spain}
\altaffiliation{Donostia International Physics Center (DIPC), P. de Manuel Lardizabal 4, 20018 San Sebasti\'{a}n, Basque Country, Spain}
\alsoaffiliation{Saint Petersburg State University, 198504 Saint Petersburg, Russia}
\author{Carlo Carbone}
\affiliation{Istituto di Struttura della Materia-CNR (ISM-CNR), S.S. 14, km 163.5, I-34149 Trieste, Italy}
\author{Roberto Flammini}
    \email{Roberto.Flammini@cnr.it}
	\affiliation{Istituto di Struttura della Materia-CNR (ISM-CNR), Via del Fosso del Cavaliere, 00133 Roma, Italy
}%

\title[]
{Temperature driven phase transition at the 
\\antimonene/\BS\ van der Waals heterostructure}


\begin{document}

\begin{tocentry}

Some journals require a graphical entry for the Table of Contents.
This should be laid out ``print ready'' so that the sizing of the
text is correct.

Inside the \texttt{tocentry} environment, the font used is Helvetica
8\,pt, as required by \emph{Journal of the American Chemical
Society}.

The surrounding frame is 9\,cm by 3.5\,cm, which is the maximum
permitted for  \emph{Journal of the American Chemical Society}
graphical table of content entries. The box will not resize if the
content is too big: instead it will overflow the edge of the box.

This box and the associated title will always be printed on a
separate page at the end of the document.

\end{tocentry}

\begin{abstract}
We report the discovery of a temperature-induced phase transition between the $\alpha$ and $ \beta$ structures of antimonene.
When antimony is deposited at room temperature on bismuth selenide, it forms domains of $\alpha$-antimonene having different orientations with respect to the substrate. 
During a mild annealing, the $\beta$ phase grows and prevails over the $\alpha$ phase, eventually 
forming a single domain that perfectly matches the surface lattice structure of bismuth selenide. 
First principles thermodynamics calculations of this van der Waals heterostructure explain the different temperature-dependent stability of the two phases and reveal a minimum energy transition path.
Although the formation energies of free-standing $\alpha$- and $\beta$-antimonene only slightly differ, the $\beta$ phase is ultimately favoured in the annealed heterostructure due to an increased interaction with the substrate mediated by the perfect lattice match. 

\end{abstract}


\section{Introduction}

Among the known allotropes of two-dimensional (2D) antimony, $\beta$-antimonene (hereafter, \bSb), defined as the buckled Sb analogue of graphene, has drawn the attention of the scientific community due to its peculiar electronic band structure which might be exploited in electronics \cite{PizziNC2016,ChangNanoS2018}, spintronics \cite{SeoN2010}, optoelectronics\cite{TsaiCC2016} and even thermal energy conversion \cite{XIE_NE_2017}. Indeed, its electronic structure is predicted to radically change depending on the thickness of the film \cite{ZhangPRB2012}. As a consequence, the electronic gap might be tuned to be negative, zero and positive and also show an indirect-to-direct transition under biaxial strain \cite{HuangNJP2014,ZhaoSR2015,Zhang2015}. 
Beside the potential of \bSb, a second 2D allotrope (the antimony equivalent of black phosphorus, hereafter \aSb) is currently attracting much interest. 
\aSb\ shows a direct instead of an indirect gap as found in $\beta$-Sb. Furthermore, the estimated electronic gap of the \ga\ phase is much lower than that of the \gb\ allotrope, although its value is still under debate \cite{Akturk2015,WangAMI2015,Zhang2016a,Wang2018}. \aSb\ also exhibits a high carrier mobility \cite{Zhang2016a,ShiAM2019} and is predicted to be an excellent absorber in the visible range \cite{Xu2017}. 

Such a wealth of properties needs to be preserved when dealing with a supporting substrate. 
Recently, van der Waals (vdW) epitaxy \cite{KOMA_ME_1984} has been suggested as a valid solution.
As a result, the number of reports of interfaces made from stable \aSb\ and \bSb\ layers on vdW substrates continues to grow .\cite{Chuang2013a, Yao2013, Ji2016, Gibaja2016, Kim2016, Wu2017, Kim2017, Fortin-Deschenes2017, Markl2017} 
Much less has been done, however, to try to control and possibly induce a structural transition from one allotrope to the other. For instance, driving mechanisms  such as electrostatic doping, molecular adsorption and the use of graphene  as a supporting  substrate \cite{Wang2018} were proposed, but no experimental counterpart has been reported. 

Here we present a method to induce the \aSb\ to \bSb\ phase transition on a bismuth selenide (\BS) monocrystal by means of annealing. 
The phase transition at the atomic scale is observed via scanning tunneling microscopy (STM), while density functional theory (DFT) calculations comprehensively explain why the \ga\ phase is energetically favoured upon growth on \BS\ at room temperature (RT) as well as why, and indeed how, the \gb\ phase forms upon gentle annealing. 

This study sheds light on the structural stability and structural phase transition undergone by the class of materials made of group 15 elements that possess $\alpha$ and $\beta$ stable allotropic phases, i.e. phosphorene, arsenene and bismuthene, all of which are of great current interest.\cite{Zhang2016a}
Furthermore, from a more applicative point of view, this mild temperature annealing method used to convert one phase to the other is fully compatible with modern electronics and may be relevant for the next generation of electronic
devices based on three-dimensional van
der Waals heterostructures.\cite{GeimN2013,AjayanPT2016,Liu2019}

The manuscript is organized as follows. 
Using STM and DFT, we identify and characterize 
the two antimonene phases and show evidence for the phase transition. 
Following this, we provide an \textit{ab initio} thermodynamics analysis of antimonene stability on \BS\ during the growth and annealing stages.
A transition path between the two phases is proposed, and the critical role played by lattice matching to the substrate is highlighted. 
Finally, the significance of this study in the design of other group 15 vdW heterostructures is discussed.

\begin{figure}[t]
    \includegraphics[width=7.8cm]{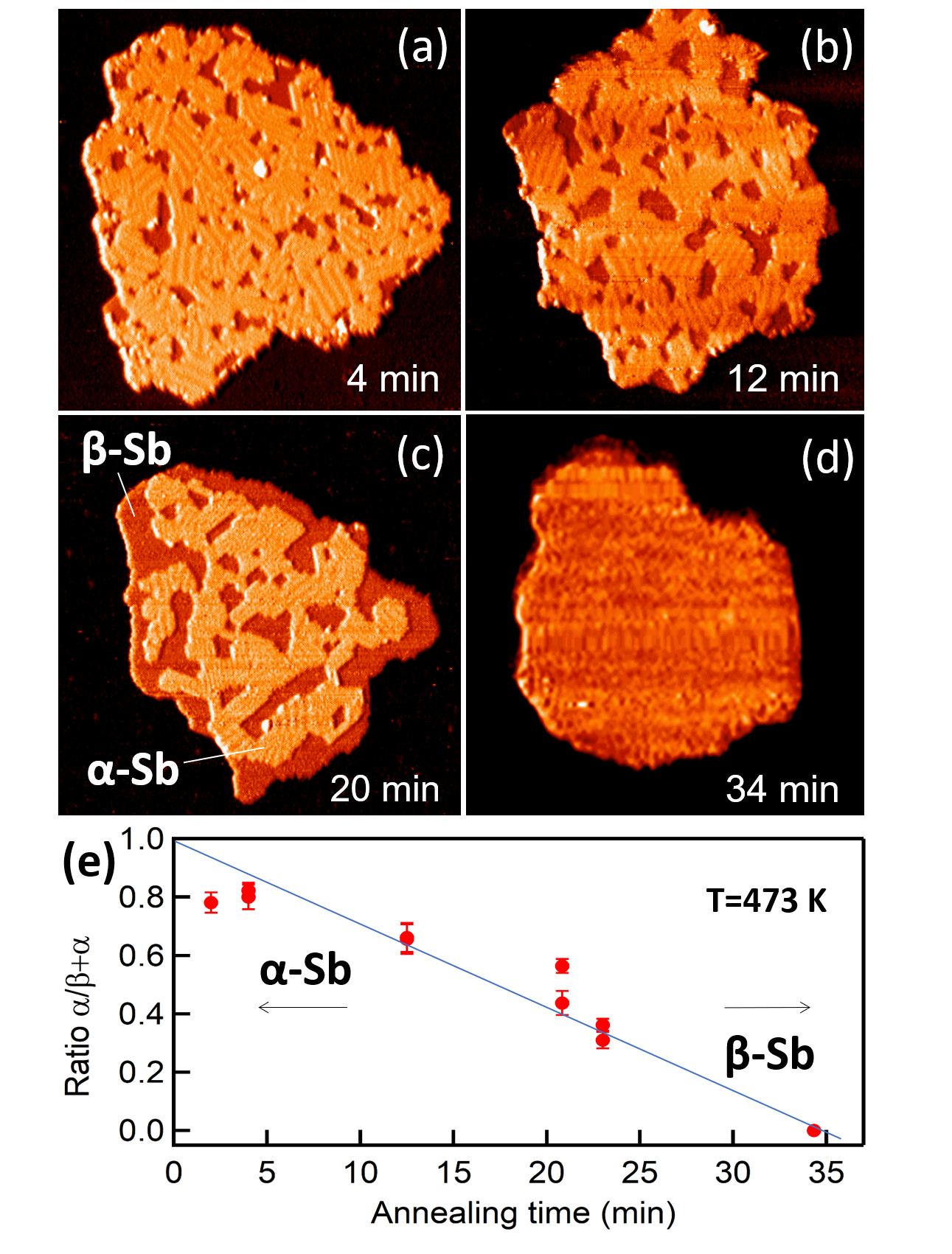}
    \caption{(Color online) Panels (a-d): Sb islands grown at RT on \BS\ and subsequently annealed at 473K for different duration time. The STM images of 100$\times$100 nm$ ^{2}$ size, have been taken at +1 V, 2 nA. 
    Bright and dark orange areas correspond to $\alpha$- and $\beta$-Sb, respectively. 
    The moir\'{e} structures seen in the orange areas are due to the superposition of a rectangular lattice ($\alpha$-Sb) on a hexagonal one (\BS). 
    (e) Ratio of $\alpha$ to $\beta$ phase as a function of the annealing time. A differentiation filter is applied to the STM images along the horizontal axes.
    \label{fig:S1}
    }
\end{figure}

\begin{figure}[tbh!]
    \includegraphics[angle=-90,width=0.45\textwidth]{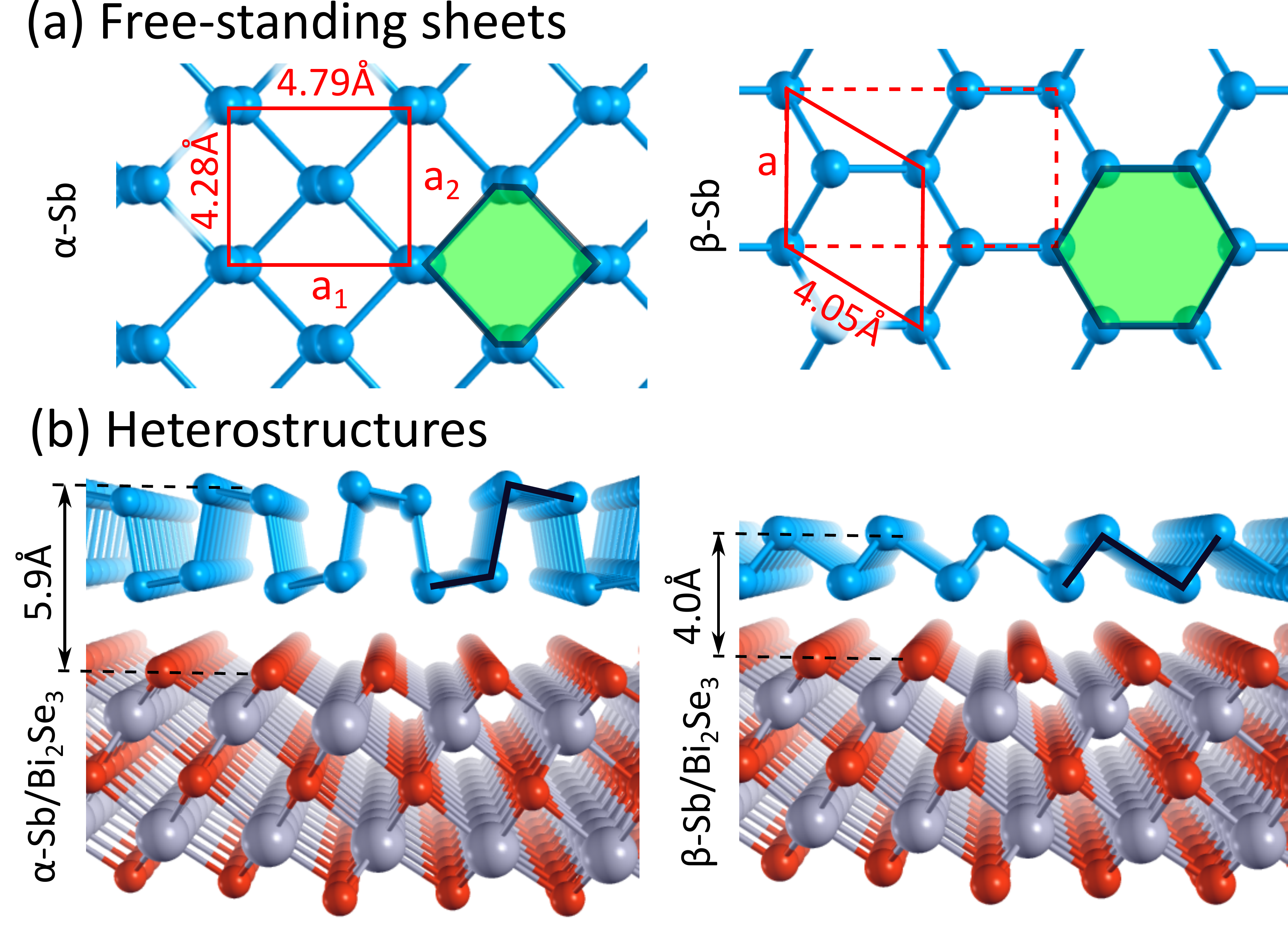}
    \caption{(a) Ball and stick models of free-standing \aSb\ and \bSb\ (top views)
    Unit cells and theoretical lattice parameters are indicated. Dashed lines indicate the four-atom \gb\ simulation cell; green shaded areas indicate the hexagonal structure common to both allotropes.
    (b) Adsorbed layers on \BS\ (only first quintiple layer shown). 
    \label{fig:models}
    }
\end{figure}

\begin{figure*}[t]
    \includegraphics[width=0.9\textwidth]{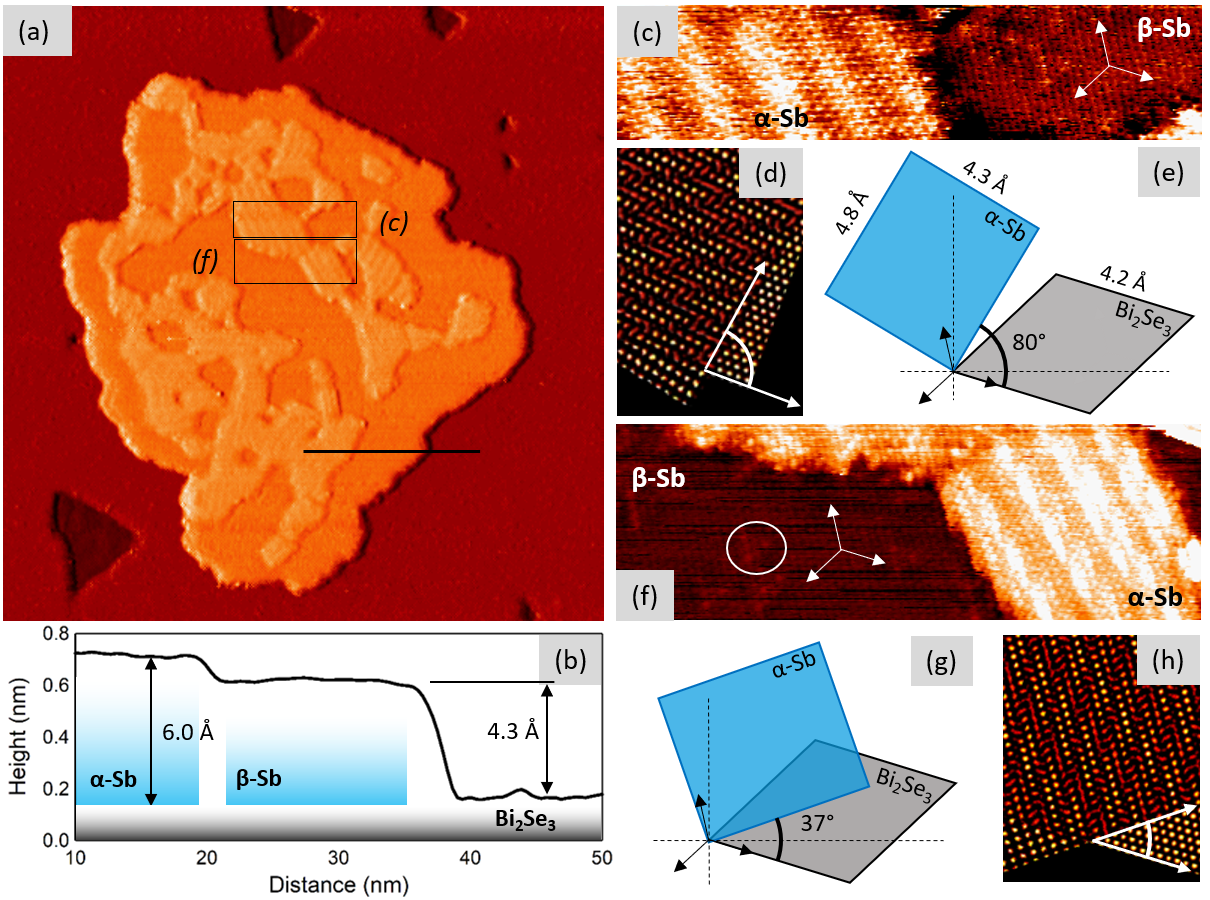}%
    \caption{(Color online) (a) STM image of an Sb island after annealing at 473K for 23 min (100$\times$100 nm$ ^{2} $, taken at +1 V, 2 nA). Brown, orange, and bright orange areas correspond to \BS\ substrate, \aSb, and \bSb, respectively.
    The black line indicates the location of the profile shown in panel (b).
    The highlighted \textit{(c)} and \textit{(f)} areas of panel (a) are magnified in the corresponding panels, where the atom-resolved \ga- and \bSb\ lattices are displayed.
    Images (c) and (f) of 25$\times$5.9 nm$ ^{2} $ and 25$\times$8.5 nm$ ^{2} $, respectively, have been taken both at +0.1 V, 20 nA.  Panels (d) and (h) show simulated moir\'{e} patterns and relative rotation angles between the two lattices. 
    Panels (e) and (g) show measured lattice parameters and demonstrate the relation between the unit cells of the \aSb\ and \BS\ lattices as used in the moir\'{e} simulation.
    \label{fig:moir}
    }
\end{figure*}

\section{Characterization of \ga\ and \gb\ antimonene phases}
When Sb is deposited on bismuth selenide at RT, islands of different size and shape form \cite{Kim2017a,Flammini2018}.
In Fig.~\ref{fig:S1}(a) we show a typical island after 4 minutes of annealing. The surface reveals tiny features that we attribute to moir\'{e} structures originating from the superposition of the rectangular lattice of the \aSb\ phase and the hexagonal lattice of the \BS\ sample surface. Apart from the moir\'{e}, it is possible to see darker areas with no stripes. 
These zones indicate the presence of \bSb: indeed, the superposition of buckled honeycomb and hexagonal structures do not generate any moir\'{e} pattern because of the perfect match between adsorbate and substrate lattices.  
In addition to the \ga\ and \gb\ structures, small protruding areas (white regions in the STM images reported in Fig.~\ref{fig:S1})  are present on the islands and are attributed to thicker Sb layers. 
Note that other 2D Sb allotropes (notably $\gamma$ and $\delta$) are predicted to be unstable \cite{Wang2015} and have never been reported in experiments. Moreover, the unbuckled \ga\ configuration features a higher total energy and has been shown to be dynamically unstable.\cite{Akturk2015}.

In Fig.~\ref{fig:S1}(b) and (c) we show two islands after 12 and 20 mins of annealing. The moir\'{e} patterns are clearly visible although the $\beta$ phase starts to be the prevailing phase. By employing a masking procedure provided by Gwyddion \cite{Gwyddion2012}, we extracted the surface ratio of the $\alpha$ to the ($\alpha$+$\beta$) phase as a function of the annealing time, as shown in Fig.~\ref{fig:S1}(e). The complement to the surface area covered by the $\alpha$ phase corresponds to the $\beta$ phase. After 34 mins the surface features the sole $ \beta$ phase (see Fig.~\ref{fig:S1}(d)). The blue straight line represents the simulated behaviour of the $\alpha$ to the ($\alpha$+$\beta$) ratio as if the phase transition was perfectly linear as a function of the temperature. 
We qualitatively observed that the higher the annealing temperature, the faster the phase transition.

In order to ascertain the presence of the \aSb\ and \bSb\ phases, we performed first principles DFT simulations of free-standing antimonene (Fig.~\ref{fig:models}(a)) and antimonene/bismuth selenide heterostructure (Fig.~\ref{fig:models}(b)). 
The $\beta$ phase has a buckled hexagonal form (point group D\textsubscript{3d}). Bulk Sb can be understood as an ABC stacking of $\beta$ monolayers.\cite{Akturk2015} 
The $\alpha$ phase has a rectangular buckled washboard structure (point group C\textsubscript{3v}). 
Our computed lattice constants for freestanding antimonene ($a=4.05$\AA\ for \gb; $a_1=4.79$\AA, $a_2=4.28$\AA\ for \ga) 
are consistent with previous works\cite{Akturk2015,Wang2015,Zhang2015} that include van der Waals interactions (the latter are known to be crucial for describing geometries of puckered monolayers\cite{Tristant2018}).
Adsorption of \aSb\ and \bSb\ on \BS\ 
was simulated using suitably oriented large supercells that minimise artificial stresses. 
The incommensurate \ga\ phase has a step height of 5.9\AA.
\bSb\ was assumed to form a commensurate adlayer.
Its mismatch with bulk \BS\ ($a_0=4.143$\AA) is 2.2\% and the computed step height 
of 4.0\AA\ is considerable smaller than that of \aSb. 
These predicted step heights and lattice parameters can thus be directly compared with data extrapolated from the STM measurements.

In Fig.~\ref{fig:moir}(a), we display a typical STM image of an island after 23 min annealing time.
The interface consists of single sheets of Sb. We traced a profile  through the island along the horizontal black line indicated in Fig.~\ref{fig:moir}(a), and show the result in Fig.~\ref{fig:moir}(b).  
Measured \emph{apparent} heights of about 6.0 and 4.3 \AA\ were extracted for the \ga\ and \gb\ phases, respectively. These data are in fair agreement with the step heights obtained in our DFT calculations, and are consistent with computed step height differences of 1.5 \AA\ between free-standing \ga\ and \gb\ reported elsewhere.\cite{Wang2015,Xu2017}
Large variations in the measured single sheet step height of \bSb\ have been reported, including 
4.5\AA\ on Bi$ _{2} $Te$ _{2} $Se\cite{Kim2017}, 
2.8\AA\ on PdTe$ _{2} $ \cite{Wu2017}, and
2.66\AA\ on Ge(111) \cite{Fortin-Deschenes2017}. 
Interlayer distances of around 6.16\AA\ have also been reported for the \ga\ phase.\cite{Wang2015}. 
Such large variations are generally attributed to the different supporting substrates, which can show large differences in terms of the local density of states, particularly if quantum spin Hall effects are taken into consideration \cite{Kim2016,Markl2017}. 
Moreover, the discrepancies could also be attributed to the well known dependency of the measured step heights on the experimental parameters chosen to carry out the STM measurements \cite{Flammini2018}.



Panels (c) and (f) of Fig. \ref{fig:moir} show atomically resolved images of two areas of panel (a). On the left side of panel (c), the moir\'{e} structure is imaged, while on the right side the typical honeycomb lattice of the \bSb\ phase is observed. The high symmetry directions of the \BS\ basal plane are indicated by white arrows. The measured lattice parameters for \aSb\ ($a_1=4.8$\AA, $a_2=4.3$\AA) are consistent with the computed values for the freestanding sheet 
[Fig.\ref{fig:models}(a)], 
indicating that \aSb\ grows as a strain-free layer on \BS.
In contrast, the measured value of $a=4.2$\AA\ for the \gb\ phase is larger than that computed for free-standing \bSb. Instead, it approaches that of bulk \BS\ (4.143\AA), in further confirmation of the commensurate growth. 

By applying a band pass FFT (Fast Fourier Transform) filter selecting the 1$ ^{st} $ order spots (separately on the left and right part of the image of panel (c)), we extracted angles relative to the STM fast scanning axis (horizontal dashed line of panel (e)). We therefore simulated the moir\'{e} patterns of panel (d) by overlaying the two lattice grids for several relative angles by employing a MatLab routine embedded in Labview.
By applying the same procedure to the STM image of panel (f), we generated the moir\'{e} structure of panel (h). The simulated images of panels (d) and (h), therefore are the result of a rotation of nearly 80$^{\circ}$ and 37$^{\circ}$ between the two unit cells, displaying a moir\'{e} period of 20 \AA\ and 27 \AA, respectively, nicely matching the experimental finding of Fig.\ref{fig:moir}(c) and (f). 
A movie of all possible mutual orientations generating moir\'{e} patterns can be found in the Supplementary Material. 

No moir\'{e} patterns are observed where the honeycomb structure of \bSb\ is present.
However, a closer look at the images of panel (c) and (f) reveals the presence of triangular defects (white circle) typical of the \BS\ surface, attributed to resonant states of Bi vacancies \cite{UrazhdinPRB2002, AlpichshevPRL2012,DaiPRL2016}. The fact that these defects are visible through the adsorbate layer can be rationalised by considering that the local density of states can be affected by defects or impurities buried deeply in the bulk of the substrate and whose trapped electronic wave functions can extend for several layers towards the surface \cite{FeenstraPRL1993}. One can argue that this evidence should be shared with the \aSb\ part of the surface. As known from the literature, free-standing \bSb\ shows a much larger gap with respect to \aSb\ \cite{Wang2015,Zhang2016a,Pumera2017a}, which allows defect states of \BS\ to be detected. Note also that the orientation of the triangular defects demonstrates the perfect alignment of the substrate and \bSb\ lattice structures.

\section{Thermodynamics of the Sb/\BS\ heterostructure}
Our calculations predict that freestanding \aSb\ has a larger cohesive energy than \bSb\ by 34 meV/atom, consistent with previous works that include non-bonding interactions\cite{Akturk2015,Wang2015,Zhang2015}. 
Although this seems to contradict the observed formation of the \gb\ phase after annealing, such calculations clearly neglect the interaction with the substrate.
Our use of large supercells for modelling the Sb/\BS\ heterostructures allows us to compute the formation energies of the incommensurate \ga\ and lattice matched \gb\ phases as well as that of (ficticious) unaligned or rotated \gbs\ phases, giving access both to energy gain on adsorption (through non-bonding interactions) as well as energy loss from strain (in matching with the substrate).

\begin{figure*}[!tbh]
    \centering
    \includegraphics[angle=-90,width=0.98\textwidth]{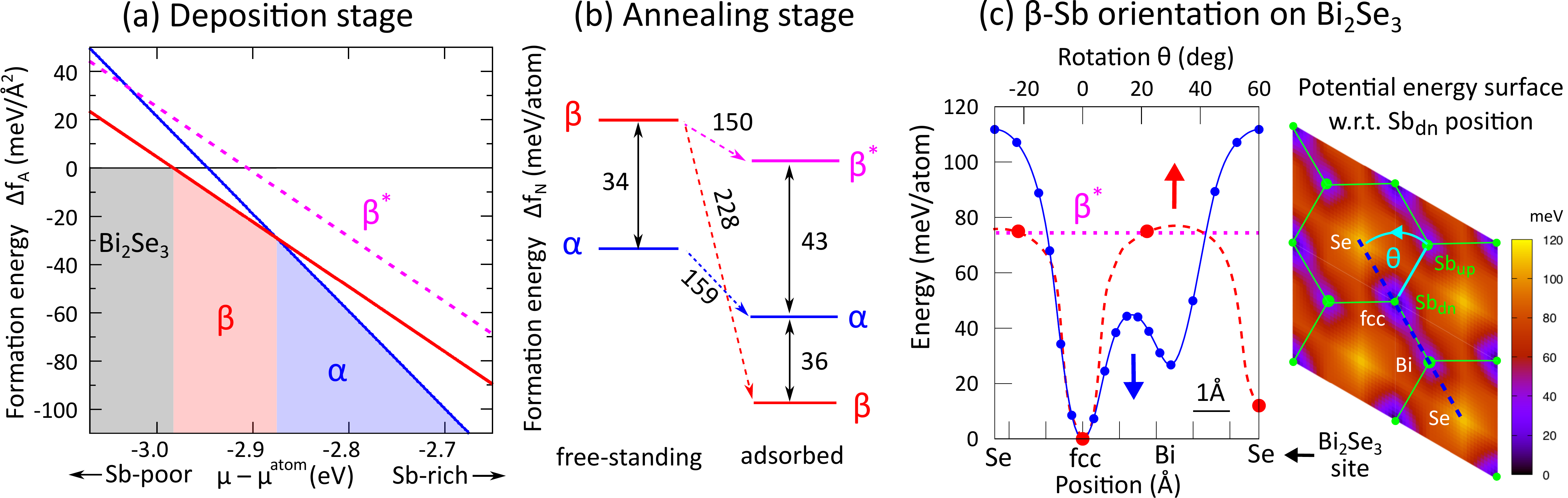}
    \caption{(a) Phase diagram during the deposition stage, showing the relative formation energy per area $\Delta f_A(\mu)$ of antimonene allotropes 
    with respect to the Sb chemical potential.
    (b) Relative formation energies per atom $\Delta f_N$ during the annealing stage, for free-standing and adsorbed sheets.
    (c) Potential energy surface (PES) for \bSb\ adsorption on \BS\ with respect to position of the lower atom Sb$_\textrm{dn}$ (right), also shown as a profile along the Se--Se path as indicated (left, solid blue line). 
    The energy dependence with respect to the sheet rotation angle about the Sb$_\textrm{dn}$ atom is estimated (left, dashed red line). 
    \label{fig:TE1}
    }
\end{figure*}

We thus employed first-principles thermodynamics based on DFT total energy calculations to explain both the initial formation of \aSb\ on \BS\ and the 
formation of the \gb\ phase after annealing.
The crucial point here is that the deposition and annealing stages obey different boundary conditions, and hence the formation energies are described by different thermodynamical potentials.
During Sb deposition, the system is described as a substrate of constant area $A$ interacting with a gaseous reservoir of Sb atoms.
The formation energy is defined \emph{per area} and obtained from the grand (Landau) potential as $f_A(\mu) = (U-TS-\mu N)/A$.
The variable $N$ is the number of adsorbed Sb atoms, $U$ is the internal energy (DFT total energy) and $S$ the entropy; the temperature $T$ is kept constant. 
The Sb chemical potential $\mu$ is also constant (but unknown), and is determined by the amount of Sb in the vapour. 
During the annealing stage, however, the number of Sb atoms $N$ is a  constant, being supplied solely by the deposited Sb adlayer. 
Here, the area of the Sb film $A$ is a free parameter because clean substrate is available between the (predominantly) \aSb\ islands. 
Now, the formation energy is defined \emph{per atom}, and obtained from the free energy as $f_N = (U-TS)/N$.
As an approximation, the entropic contribution $TS$ of the considered adsorbate models are taken to be equal.\cite{Sanna2010} Recent studies of phosphorene adsorbed on gold have shown that the entropic contribution to the energy difference between its \gb\ and \ga\ phases is only about 10meV/atom at RT.\cite{Tristant2018}


Fig.~\ref{fig:TE1}(a) shows the phase diagram corresponding to the deposition stage. Formation energies of adsorbed \ga\ and \gb\ phases are reported relative to clean \BS; 
$\mu$ is given relative to that of atomic Sb.
For Sb-rich conditions, which correspond to the usual experimental growth condition, the \aSb\ phase is the most stable.
The commensurate \bSb\ phase is stable within a narrow range of $\mu$, explaining why small amounts of \bSb\ are also found during growth.  
Unaligned \gbs\ phases are never stable, consistent with the experimental observation of only one \gb\ phase alignment across the whole surface.
The presence of the gaseous Sb reservoir is, therefore, the determining factor during deposition, favouring phases that pack more atoms into the same area (steeper slope in Fig.~\ref{fig:TE1}).
During the subsequent annealing stage, the formation energy per atom $f_N$ determines the relative stability of the phases. 
Cohesive, strain, vdW, and adsorption (binding) energies all contribute to $f_N$.  
It is therefore instructive to compare formation energies of adsorbed phases with the freestanding ones as shown in Fig.~\ref{fig:TE1}(b). 
The difference in $f_N$ between free-standing \ga\ and \gb\ sheets is 34 meV/atom, with \ga\ the more stable. 
%
When the interaction with the substrate is included, however, the lattice-matched \gb\ phase becomes more stable than the incommensurate \ga\ phase by 36 meV/atom (i.e. a relative gain of 70 meV/atom). 
In the absence of the Sb reservoir, therefore, the \bSb\ phase is ultimately the most favoured. Experimentally, it is obtained after light annealing. 

Fig.~\ref{fig:TE1}(b) also demonstrates that the unaligned \gbs\ phase  is less stable than both adsorbed \ga\ and \gb\ phases, even though \gbs\ is a strain free adlayer.
In fact, \gb\ is stable only for a unique orientation and
alignment on the \BS\ surface.
Fig.~\ref{fig:TE1}(c) shows the potential energy surface (right) and its line profile (left, solid blue line) with respect to the \bSb\ sheet position, as well as the energy dependence on sheet rotation angle (dashed red line) estimated from a few computed datapoints.
Laterally shifting or rotating the \gb\ sheet even by small amounts imposes a sharp energy penalty, with local Sb-on-Se configurations particularly punitive. 
Such repulsive interactions cause any randomly oriented sheet to move an extra $\sim0.7$~\AA\ away from the substrate: indeed, \gbs\ and \ga\ phases both lie about 3.1\AA\ from the surface, in comparison to 2.4\AA\ for the aligned \gb\ phase. 
Lattice matching thus allows the antimonene sheet to move closer to the substrate while the perfect atomic alignment maximizes favourable non-bonding interactions (vdW forces are short-ranged and additive). Of the 228~meV/atom gained on adsorption, 181~meV (80\%) comes from the vdW interactions. 
Last, it is interesting to note that spin-orbit coupling accounts for a considerable 36~meV/atom of the \gb\ adsorption energy, which translates into 41\% of the formation energy difference between \gb\ and \ga\ phases (see Supplementary Material for details).

\section{VdW-mediated \aSb$\to$\bSb\ transition}

\begin{figure*}[!tbh]
    \centering
    \includegraphics[angle=-90,width=1\textwidth]{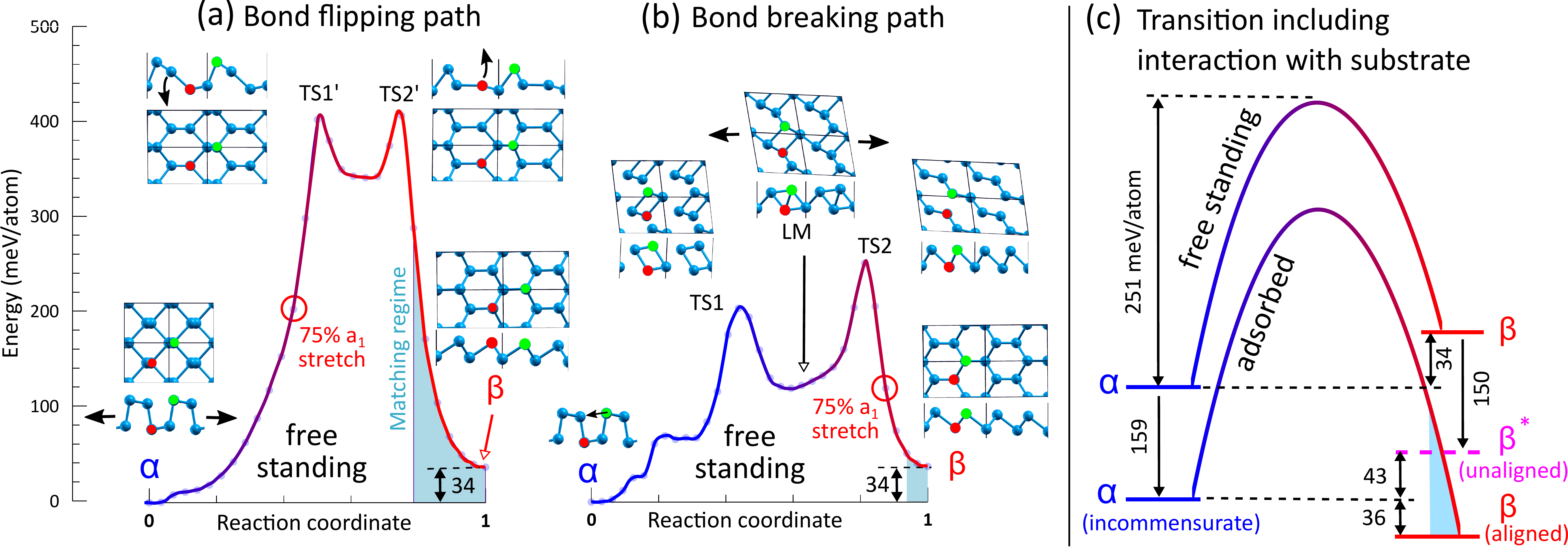}
    \caption{Transition pathways from \ga\ to \gb\ phases of freestanding antimonene. 
    (a) In the bond flipping path, rapid cell extension along $a_1$ is followed by stepwise flipping of bonds. The open circle indicates the point at which the cell parameter $a_1$ reaches 75\% of its final value (in the \gb\ phase).
    Shaded (blue) areas indicate where antimonene unit cells are at least 95\% lattice matched with \BS.
    (b) The bond breaking path involves cell distortion and multiple breaking/formation of bonds; cell expansion occurs later in the pathway. 
    (c) Schematic of phase transition including stabilizing effect of substrate. Energies in meV/atom.
    \label{fig:barrier}
    }
\end{figure*}

Our calculations thus fully support the STM observations of these two distinct phases. Nonetheless, transformation from the \ga\ to the \gb\ phase must involve a barrier, as evidenced by the need for the low temperature annealing step. 
Simulation of the transition pathway is made difficult by the presence of the substrate and by the different cell sizes of \ga\ and \gb\ phases. 
We thus adopted the variable-cell nudged elastic band (vcNEB) scheme to model the transition for the free-standing layers, and account for the  substrate afterwards. Similar approaches have recently been applied to model the free standing \ga$\rightarrow$\gb\ transition in phosphorene\cite{Tristant2018} and antimonene\cite{Wang2018}, albeit in a restricted manner. 


Two possible kinds of pathways were identified as shown in Fig.~\ref{fig:barrier}(a) and (b).
The `bond flipping pathway' (a) describes a displacive process whereby some atoms in the upper atomic layer of the Sb sheet move to the lower atomic layer; no bonds are broken. 
Since chains of buckled Sb hexagons (green shaded areas in Fig.~\ref{fig:models}(a)) are common to both phases, a direct \ga$\rightarrow$\gb\ transformation can occur simply by flipping the bonds linking the hexagons.
To facilitate this process, the \ga\ cell first rapidly expands through stretching of $a_1$ (the point where it reaches 75\% of its final value in the \gb\ phase is indicated). 
The two equal height barriers (about 400 meV/atom) correspond to transition states in which each Sb atom in the flipping bond successively experiences a flat $sp^2$-like geometry (see TS1' and TS2' in Fig.~\ref{fig:barrier}(a)). This stepwise process predicts a higher barrier than that reported elsewhere (250 eV/atom) for a single step pathway of hole-doped antimonene,\cite{Wang2018} but is similar to that predicted for the transition from black to blue phosphorene (480 meV/atom).\cite{Tristant2018} 


A lower energy transition pathway  was identified, however, which is far more complex.
This asymmetric `bond breaking pathway', shown in Fig.~\ref{fig:barrier}(b), is a reconstructive process that involves breaking and formation of multiple Sb--Sb bonds.
Interestingly, the local minimum (LM) between the two barriers of unequal height corresponds to a structure very similar to the metastable $\eta$ phase reported by Zhang et al.\cite{Zhang2016a} 
In contrast to the previous pathway, however, Sb atoms remain in the \emph{same} atomic plane throughout, instead undergoing lateral translations until the sheet acquires the hexagonal geometry of the \gb\ phase.
During the initial stages of the transition the unit cell undergoes considerable shear distortion; the volume remains fairly constant.
Cell expansion occurs mostly in the latter stages, as indicated in Fig.~\ref{fig:barrier}(b).
The minimum barrier for the freestanding \ga$\to$\gb\ transformation was thus identified at 251~meV/atom. 
Note that precise pathways and transition barriers can only be identified via algorithms that consider atomic forces and cell stresses on a free and equal footing.\cite{Qian2013,Sheppard2012a} 
Our computed barrier height is considerably lower than that reported (650meV/atom) for a similar pathway in phosphorene.\cite{Tristant2018} The difference may be attributed to the 40\% smaller bond dissociation energy of antimony, which allows traditionally unfavoured paths (involving breaking of chemical bonds) to be preferred over alternatives that purely feature bond rotation. 

Finally, we consider the influence of the \BS\ substrate on the transition path.
The lowest energy path identified for free-standing antimonene is schematically depicted by the upper line in Fig.~\ref{fig:barrier}(c). 
As discussed earlier, adsorption on \BS\ reorders the \ga\ and \gb\ formation energies such that the latter becomes more stable by 36 meV.
One might therefore expect the pathway for the adsorbed layers (lower curve in Fig.~\ref{fig:barrier}(c)) to follow this trend and exhibit a lower barrier height. 
However, this is not the case.
Formation of the stable (aligned) \bSb\ adlayer can only occur if the antimonene cell is reasonably lattice matched to the surface cell of \BS. 
Our vcNEB calculations indicate that a $<5\%$ mismatch occurs only near the end of the pathway (shaded blue areas in Fig.~\ref{fig:barrier}). 
Up until this point, therefore, the transition pathway of the \emph{adsorbed} layer can be understood as moving from the incommensurate \ga\ phase to the unaligned \gbs\ phase. 
As the \ga\ and \gbs\ adsorption energies differ by only 9 meV, it is reasonable to infer that all intermediate states are similarly weakly bound to, and equidistant ($\sim3.1$\AA) from, the \BS\ surface.
The barrier profile and height for this stage of the transition should thus remain \emph{similar} to that of the free-standing layers.
Once the antimonene layer enters the matching regime, however, it 
can then rigidly shift or rotate in-plane until a commensurate alignment is found. At this point, the stable \gb\ phase forms: the antimonene layer moves 0.7\AA\ closer to the substrate and rapidly gains a further 79 meV/atom, as discussed in Fig.~\ref{fig:TE1}. 


Although this two-stage transition process through an intermediate \gbs\ phase is unlikely to occur on a large scale in the real system, it may happen at the edge of the Sb islands on a scale of several unit cells, perhaps acting as a seed for further antimonene alignment. 
Indeed, our STM images typically show that regions of \gb\ phase develop mostly around the island edges.
It is also notable that the predicted barrier heights for forward and backward transitions are thus quite similar (251 vs. 287 meV/atom).
This is in apparent contradiction with the experimental observation with STM of a 100\% transition to the \gb\ phase after annealing.
However, we recall that the \gb\ phase lies in a quite steep potential well with respect to translation or rotation of the Sb sheet (Fig.~\ref{fig:TE1}(c)). In order for the reverse transition to occur, the \gb\ sheet must therefore first escape the well via vertical displacement away from the substrate before any \emph{lateral} atomic motion or cell distortion can occur---i.e., it must recover the \gbs\ phase. At this point the (barrierless) \gbs$\to$\gb\ transition is far more likely to occur than the reverse \gbs$\to$\ga\ process. 
Ultimately, the unique lattice matching of the \gb\ phase on \BS\ means any transformation back to the \ga\ phase is \emph{kinetically} limited, in agreement with the experimental observation.

\section{Conclusions}

VdW epitaxy is achieved when a substrate is naturally passivated, allowing the adsorbate to grow with its orientation and lattice structure independent of the substrate structure \cite{KOMA_ME_1984}. Our results for both \ga\ and \gb\ phases of antimonene supported on \BS\ show a behaviour typical of an epitaxial vdW heterostructure. 
The observation of small domains of \aSb\ with no defined orientation 
indicates that the interaction with the substrate is too weak to induce any strain or preferential orientation.
As for \bSb, even if the orientation of the domains is unique, the interaction with the substrate can nonetheless be described as vdW mediated \cite{Liu2019}.
Previous core-level spectroscopy and angle resolved photoelectron spectroscopy studies demonstrated that the electronic structure can be ascribed to the sole presence of \bSb, and not to a Bi-Se-Sb alloy \cite{Jin2016,Lei2016a,Kim2017}. In other words, the interaction is strong enough to align the lattice but not so strong to be described by covalent or ionic bonds.

Our calculations demonstrate that the 100\% transformation to \bSb\ essentially occurs due to specific lattice matching of \bSb\ with \BS. 
To induce growth of the \gb\ allotrope in other 2D X-enes (where X=P,Sb,As,Bi) we therefore suggest using hexagonal symmetry substrates that offer favourable matching. 
In heterostructures where such matching does not occur, only incommensurate, weakly vdW-bound \ga\ and \gb\ phases may form. 
During growth of such systems, the \ga\ phase adlayer should remain favoured due to its higher atomic density. 
Annealing is unlikely to induce a complete transition to the \gb\ phase, however. Post-anneal islands may feature a coexistence of both \ga\ and unaligned \gb\ phases, depending on the specific barrier height. While our calculations showed that a reconstructive (bond-breaking) phase transition is favoured in 2D antimony, a displacive phase transition may be favoured for other elemental species like P,\cite{Tristant2018}  thus limiting the available pathways and imposing higher barriers.


The Sb/\BS\ interface may provide a platform for studying both allotropic forms of  antimonene at the same time. On one hand, we can exploit the phase diagram to induce formation of one allotrope with respect to the other by changing the Sb partial pressure  during growth \cite{SannaPRB2012}; on the other hand, the temperature control allows us to change their relative population. This may permit comparative studies of both phases without the drawbacks of systematic errors due to the preparation of two different samples. Moreover, a more profound understanding of the class of materials made by an interface comprising a stable antimonene allotrope and a vdW topological insulator can be reached.\cite{Xia2014,Kim2017,Jin2016} Interpretation of the electronic structure in such heterostructures is challenging and often affected by the absence of a single structural phase. 

The novel combination of \textit{ab initio} thermodynamics and transition path simulations presented here thus opens new possibilities for designing and engineering of other vdW bound heterostructures. Applications aiming at the exploitation of topology--protected electronic states may indeed benefit from the fine tuning of the adsorbate-substrate structure symmetries \cite{Tong2016}.


In summary, we have proposed a method to induce a structural phase transition from $\alpha$ to $\beta$-Sb by means of low temperature annealing. Theoretical calculations show that while both $\alpha$--Sb and $\beta$--Sb are loosely bound to the substrate, the latter offers increased thermodynamical stability under limited Sb conditions and gives rise to a phase exhibiting an almost perfect lattice match and orientation with the substrate.

\section{Methods}
\subsection{Sample preparation and STM measurements}

The \BS\ samples are (0001)-oriented single crystals grown
by the modified Bridgman method. These samples have been
previously characterised by x-ray diffraction. The \BS\
surface was prepared by cleavage in-situ, in an ultra high
vacuum (UHV) chamber. The structural ordering of the sample was confirmed by low energy electron diffraction (LEED) and by scanning tunnelling Microscopy (STM). 

Antimony was sublimated from a temperature-controlled Knudsen cell. All depositions were performed at RT with a deposition rate of ~0.5 \AA/min. 1 ML of Sb on \BS\ corresponds to surface density of 0.033 \AA$ ^{-2} $, that is, to an equivalent thickness of 2.03 \AA. The reported STM images where collected after deposition of about 1 ML of Sb, corresponding to half the full surface coverage. 

The temperature calibration was done by exploiting the ($2\sqrt{3}\times2\sqrt{3}~ R30^{\circ}$)$\rightarrow$($\sqrt{3}\times\sqrt{3}~R30^{\circ}$) transition occurring at 473 K for more than 1.5 ML of Sb deposited on Ag(111) \cite{Noakes1997a}.   

STM images were recorded by using an Omicron LT-STM housed in a UHV vacuum chamber with base pressure below 1$\times$10$^{-10} $ mbar. The STM images were acquired at 80K using a W tip cleaned by electron bombardment in UHV. The STM scanner was calibrated measuring the clean \BS(0001) surface. Bias voltage is referred to sample, hence positive (negative) bias corresponds to empty (filled) states. STM measurements have been performed with a surface coverage of about 30\% of Sb/\BS.

\subsection{Calculation method}

Density functional theory (DFT) calculations were performed using the quantum-ESPRESSO code.\cite{Giannozzi2009}
We used the PBE exchange-correlation functional\cite{Perdew1996} 
along with a semi-empirical van der Waals correction (D2)\cite{Grimme2006} and optimized, fully-relativistic norm-conserving pseudopotentials (cutoff 85 Ry).\cite{VanSetten2018,Hamann2013}
PBE-D2 was found to yield good agreement for the experimental lattice parameters of bulk Sb and was thus adopted throughout (see Supplementary Material for full details). 

Structural optimizations were performed until forces were below 10meV/\AA\ and stresses below 0.2 Kbar. Symmetry was not enforced.  

Adsorption energies of \ga\ and \gb-Sb on hexagonal \BS\ were computed using a $(1\times4$) \BS\ supercell. 
With this setup, the commensurate $(1\times4$) $\beta$-Sb cell has a mismatch of 2.2\%, while the incommensurate $(1\times3)$ $\alpha$-Sb cell has a mismatch of 3.3\% and 0.1\% along the $a_1$ and $a_2$ directions, respectively.
Rotated and unaligned/shifted \gb\ and \gbs\ adlayer calculations used a ($\sqrt7\times\sqrt7)R21.8^\circ$ supercell. The PES was computed in a $(1\times1)$ cell.
The $\alpha\to\beta$ phase transition in free-standing Sb-ene was simulated using the variable cell nudged elastic band scheme (vcNEB)\cite{Qian2013,Dong2013} as implemented in the USPEX code.\cite{oganov2006,oganov2011,Lyakhov2010} A four-atom cell was used and several initial paths were considered.
Spin-orbit coupling was included except in vcNEB and PES calculations.
\begin{acknowledgement}

The Deutsche Forschungs\-gemein\-schaft (DFG) is gratefully acknowledged for financial support via projects FOR1700, Project No. SA 1948/1-2. The H\"ochst\-leistungrechen\-zentrum at Stuttgart and at Berlin (HLRS, HRLN) as well as CINECA, Italy via the ISCRA initiative are gratefully acknowledged for grants of high-performance computing time and support. We acknowledge computational resources provided by the HPC Core Facility and the HRZ of the Justus-Liebig-Universit\"at Gie\ss en. The authors thank G. Emma and M. Rinaldi for their valuable technical assistance.    

\end{acknowledgement}

\begin{suppinfo}

Details of the DFT simulations: influence of exchange-correlation functional and spin orbit coupling, unit cells for modelling incommensurate adlayers, thermodynamical framework, and role of strain. 
Movie showing the development of moir{\'e} patterns for a wide range of rotation angles.

\end{suppinfo}

\bibliography{Projects-Active-Papers-Sb-ene}

\end{document}